\documentclass[aps,prd,twocolumn,showpacs,amsmath,amssymb]{revtex4-1}
\usepackage{amsmath}
\usepackage{graphicx}
\usepackage{subfigure}
\usepackage{epstopdf}
\usepackage{color}
\usepackage{multirow}
\usepackage{setspace}
\usepackage{overpic}
\usepackage{amssymb}
\usepackage[bookmarksnumbered, pdfstartview=FitH,colorlinks,urlcolor=blue, citecolor=blue,linkcolor=blue] {hyperref}
\usepackage{lineno}
\usepackage{bm}
\usepackage{rotating}
\usepackage[utf8]{inputenc}

\let\oldequation\equation
\let\oldendequation\endequation

\renewenvironment{equation}
  {\linenomathNonumbers\oldequation}
  {\oldendequation\endlinenomath}

\begin{document}
%\linenumbers

\title{\bf \boldmath
Measurements of the absolute branching fractions of $D^{0(+)}\to K\bar K\pi\pi$ decays
}

\author{
M.~Ablikim$^{1}$, M.~N.~Achasov$^{10,c}$, P.~Adlarson$^{64}$, S. ~Ahmed$^{15}$, M.~Albrecht$^{4}$, R.~Aliberti$^{28}$, A.~Amoroso$^{63A,63C}$, Q.~An$^{60,48}$, ~Anita$^{21}$, X.~H.~Bai$^{54}$
% ...
\\
\vspace{0.2cm}
(BESIII Collaboration)\\
\vspace{0.2cm} {\it
$^{1}$ Institute of High Energy Physics, Beijing 100049, People's Republic of China\\
% ...
$^{4}$ Bochum Ruhr-University, D-44780 Bochum, Germany\\
% ...
$^{10}$ G.I. Budker Institute of Nuclear Physics SB RAS (BINP), Novosibirsk 630090, Russia\\
% ...
$^{15}$ Helmholtz Institute Mainz, Johann-Joachim-Becher-Weg 45, D-55099 Mainz, Germany\\
% ...
$^{21}$ Indian Institute of Technology Madras, Chennai 600036, India\\
% ...
$^{28}$ Johannes Gutenberg University of Mainz, Johann-Joachim-Becher-Weg 45, D-55099 Mainz, Germany\\
% ...
$^{48}$ State Key Laboratory of Particle Detection and Electronics, Beijing 100049, Hefei 230026, People's Republic of China\\
% ...
$^{54}$ University of Jinan, Jinan 250022, People's Republic of China\\
% ...
$^{60}$ University of Science and Technology of China, Hefei 230026, People's Republic of China\\
% ...
$^{63}$ (A)University of Turin, I-10125, Turin, Italy; (B)University of Eastern Piedmont, I-15121, Alessandria, Italy; (C)INFN, I-10125, Turin, Italy\\
$^{64}$ Uppsala University, Box 516, SE-75120 Uppsala, Sweden\\
% ...
$^{c}$ Also at the Novosibirsk State University, Novosibirsk, 630090, Russia\\
% ...
}
}
%%% Local Variables:
%%% mode: latex
%%% TeX-master: "draft_BAM186"
%%% End:

\begin{abstract}
Based on 2.93~fb$^{-1}$ $e^+e^-$ collision data taken at center-of-mass energy of 3.773 GeV by the
BESIII detector,
we report the measurements of the absolute branching fractions of $D^0\to K^+K^-\pi^0\pi^0$,
$D^0\to K^0_SK^0_S\pi^+\pi^-$,
$D^0\to K^0_SK^-\pi^+\pi^0$,
$D^0\to K^0_SK^+\pi^-\pi^0$,
$D^+\to K^+K^-\pi^+\pi^0$,
$D^+\to K^0_SK^+\pi^0\pi^0$,
$D^+\to K^0_SK^-\pi^+\pi^+$,
$D^+\to K^0_SK^+\pi^+\pi^-$, and
$D^+\to K^0_SK^0_S\pi^+\pi^0$.
The decays $D^0\to K^+K^-\pi^0\pi^0$, $D^0\to K^0_SK^-\pi^+\pi^0$, $D^0\to K^0_SK^+\pi^-\pi^0$, $D^+\to K^0_SK^0_S\pi^+\pi^0$, and $D^+\to K^0_SK^+\pi^0\pi^0$ are observed for the first time.
The branching fractions of the decays $D^0\to K^0_SK^0_S\pi^+\pi^-$, $D^+\to K^+K^-\pi^+\pi^0$, $D^+\to K^0_SK^-\pi^+\pi^+$, and $D^+\to K^0_SK^+\pi^+\pi^-$ are measured with improved precision
compared to the world-average values.
\end{abstract}

\pacs{13.20.Fc, 14.40.Lb}

\maketitle

\oddsidemargin  -0.2cm
\evensidemargin -0.2cm

\section{Introduction}
Multi-body hadronic $D^{0(+)}$ decays provide an ideal laboratory to study strong and weak interactions.
Amplitude analyses of these decays offer comprehensive information of quasi-two-body $D^{0(+)}$ decays, which are important to explore $D\bar D^0$ mixing,
charge-parity ($CP$) violation and quark SU(3)-flavor asymmetry breaking phenomenon~\cite{ref5,theory_1,theory_2,chenghy1,yufs}.
In particular, for the search of $CP$ violation, it is important to understand the intermediate structures for the singly Cabibbo-suppressed decays
of $D^{0(+)}\to K\bar K\pi\pi$~\cite{xwkang,Charles:2009ig,yufs-cpv}.

Current measurements of the $D^{0(+)}\to K\bar K\pi\pi$ decays containing $K^0_S$ or $\pi^0$ are limited~\cite{pdg2018}.
The branching fractions (BFs) of
$D^0\to K^0_SK^0_S\pi^+\pi^-$~\cite{FOCUS_kskspipi,ARGUS_kkpipi},
$D^+\to K^0_SK^-\pi^+\pi^+$~\cite{FOCUS_kskpipi},
$D^+\to K^0_SK^+\pi^+\pi^-$~\cite{FOCUS_kskpipi}, and
$D^+\to K^+K^-\pi^+\pi^0$~\cite{ACCMOR_kkpipi0} were only determined relative to some well known decays
or via topological normalization, with poor precision.
This paper presents the first direct measurements of the absolute BFs for the decays
$D^0\to K^+K^-\pi^0\pi^0$,
$D^0\to K^0_SK^0_S\pi^+\pi^-$,
$D^0\to K^0_SK^-\pi^+\pi^0$,
$D^0\to K^0_SK^+\pi^-\pi^0$,
$D^+\to K^+K^-\pi^+\pi^0$,
$D^+\to K^0_SK^+\pi^0\pi^0$,
$D^+\to K^0_SK^-\pi^+\pi^+$,
$D^+\to K^0_SK^+\pi^+\pi^-$, and
$D^+\to K^0_SK^0_S\pi^+\pi^0$.
The $D^0\to K^0_SK^0_S\pi^0\pi^0$ decay is not included since it suffers from poor statistics and high background.
Throughout this paper, charge conjugate processes are implied.
An $e^+e^-$ collision data sample corresponding to an integrated luminosity of 2.93~fb$^{-1}$~\cite{lum_bes3} collected at a
center-of-mass energy of $\sqrt s=$ 3.773~GeV  with the BESIII detector is used to perform this analysis.

\section{BESIII detector and Monte Carlo simulation}

The BESIII detector is a magnetic
spectrometer~\cite{BESIII} located at the Beijing Electron
Positron Collider (BEPCII)~\cite{Yu:IPAC2016-TUYA01}. The
cylindrical core of the BESIII detector consists of a helium-based
 multilayer drift chamber (MDC), a plastic scintillator time-of-flight
system (TOF), and a CsI(Tl) electromagnetic calorimeter (EMC),
which are all enclosed in a superconducting solenoidal magnet
providing a 1.0~T magnetic field. The solenoid is supported by an
octagonal flux-return yoke with resistive plate counter muon
identifier modules interleaved with steel. The acceptance of
charged particles and photons is 93\% over $4\pi$ solid angle. The
charged-particle momentum resolution at $1~{\rm GeV}/c$ is
$0.5\%$, and the $dE/dx$ resolution is $6\%$ for the electrons
from Bhabha scattering. The EMC measures photon energies with a
resolution of $2.5\%$ ($5\%$) at $1$~GeV in the barrel (end cap)
region. The time resolution of the TOF barrel part is 68~ps, while
that of the end cap part is 110~ps.

Simulated samples produced with the {\sc geant4}-based~\cite{geant4} Monte Carlo (MC) package including the geometric description of the BESIII detector and the
detector response, are used to determine the detection efficiency
and to estimate the backgrounds. The simulation includes the beam-energy spread and initial-state radiation (ISR) in the $e^+e^-$
annihilations modeled with the generator {\sc kkmc}~\cite{kkmc}.
The inclusive MC samples consist of the production of $D\bar{D}$
pairs with consideration of quantum coherence for all neutral $D$
modes, the non-$D\bar{D}$ decays of the $\psi(3770)$, the ISR
production of the $J/\psi$ and $\psi(3686)$ states, and the
continuum processes.
The known decay modes are modeled with {\sc
evtgen}~\cite{evtgen} using the BFs taken from the
Particle Data Group (PDG)~\cite{pdg2018}, and the remaining unknown decays
from the charmonium states are modeled with {\sc
lundcharm}~\cite{lundcharm}. The final-state radiations
from charged final-state particles are incorporated with the {\sc
photos} package~\cite{photos}.

\section{Measurement Method}

The $D^0\bar D^0$ or $D^+D^-$ pair is produced without an additional hadron in $e^+e^-$ annihilations at $\sqrt s=3.773$ GeV. This process
offers a clean environment to measure the BFs of the hadronic $D$ decay with the double-tag (DT) method.
The single-tag (ST) candidate events are selected by reconstructing a $\bar D^0$ or $D^-$ in the following hadronic final states:
$\bar D^0 \to K^+\pi^-$, $K^+\pi^-\pi^0$, and $K^+\pi^-\pi^-\pi^+$, and
$D^- \to K^{+}\pi^{-}\pi^{-}$,
$K^0_{S}\pi^{-}$, $K^{+}\pi^{-}\pi^{-}\pi^{0}$, $K^0_{S}\pi^{-}\pi^{0}$, $K^0_{S}\pi^{+}\pi^{-}\pi^{-}$,
and $K^{+}K^{-}\pi^{-}$.
The event in which a signal candidate is selected in the presence of an ST $\bar D$ meson,
is called a DT event.
The BF of the signal decay is determined by
\begin{equation}
\label{eq:br}
{\mathcal B}_{{\rm sig}} = N^{\rm net}_{\rm DT}/(N^{\rm tot}_{\rm ST}\cdot\epsilon_{{\rm sig}}),
\end{equation}
where
$N^{\rm tot}_{\rm ST}=\sum_i N_{{\rm ST}}^i$ and $N^{\rm net}_{\rm DT}$
are the total yields of the ST and DT candidates in data, respectively.
$N_{{\rm ST}}^i$ is the ST yield for the tag mode $i$. For the signal decays involving $K^0_S$ meson(s) in the final states, $N^{\rm net}_{\rm DT}$ is
the net DT yields after removing the peaking background from the corresponding non-$K^0_S$ decays. For the other signal decays, the variable corresponds to
the fitted DT yields as described later.
Here, $\epsilon_{{\rm sig}}$ is the efficiency of detecting the signal $D$ decay, averaged over the tag mode $i$, which is given by:
\begin{equation}
\label{eq:eff}
\epsilon_{{\rm sig}} = \sum_i (N^i_{{\rm ST}}\cdot\epsilon^i_{{\rm DT}}/\epsilon^i_{{\rm ST}})/N^{\rm tot}_{\rm ST},
\end{equation}
where $\epsilon^i_{{\rm ST}}$ and $\epsilon^i_{{\rm DT}}$ are the efficiencies of detecting ST and DT candidates in the tag mode $i$, respectively.

\section{Event selection}

The selection criteria of $K^\pm$, $\pi^\pm$, $K^0_S$, and $\pi^0$ are the same as those used in the analyses presented in
Refs.~\cite{epjc76,cpc40,bes3-pimuv,bes3-Dp-K1ev,bes3-etaetapi,bes3-omegamuv,bes3-etamuv,bes3-etaX}.
All charged tracks, except those from $K^0_{S}$ decays, are required to have a polar angle $\theta$ with respect to the beam direction
within the MDC acceptance $|\rm{cos\theta}|<0.93$,  and a distance of closest approach to the interaction point (IP) within 10~cm along the beam direction
and within 1~cm in the plane transverse to the beam direction. Particle identification (PID) for charged pions, kaons, and protons is performed by
exploiting TOF information and the specific ionization energy loss $dE/dx$ measured by the MDC.
The confidence levels for pion and kaon hypotheses ($CL_{\pi}$ and $CL_{K}$) are calculated. Kaon and pion candidates are required to
satisfy $CL_{K}>CL_{\pi}$ and $CL_{\pi}>CL_{K}$, respectively.

The $K^0_S$ candidates are reconstructed from two oppositely charged tracks to which no PID criteria are applied and which masses are assumed to be that of pions.
The charged tracks from the $K^0_S$  candidate must satisfy $|\rm{cos\theta}|<0.93$. In addition, due to the long lifetime of the $K^0_S$  meson,
there is a less stringent criterion on the distance of closest approach to the IP in the beam direction of less than 20~cm and no requirement on the
distance of closest approach in the plane transverse to the beam direction.
%The $K^0_S$ candidates are reconstructed via their dominant decay $K^0_S\to\pi^+\pi^-$. The charged tracks from $K^0_{S}$ decays are required to
%satisfy $|V_{z}|<$ 20\,cm, and no requirement on $|V_{xy}|$ is applied. The two oppositely charged tracks are assigned as $\pi^+\pi^-$ without PID.
Furthermore, the $\pi^+\pi^-$ pairs are constrained to originate from a common vertex and their invariant mass is required to be within $(0.486,0.510)~{\rm GeV}/c^2$,
which corresponds to about three times the fitted resolution around the nominal $K^0_S$ mass. The decay length of the $K^0_S$ candidate is required to be
larger than two standard deviations of the vertex resolution away from the IP.

The $\pi^0$ candidate is reconstructed via its $\gamma\gamma$ decay. The photon candidates are selected using the information from the EMC shower.
It is required that each EMC shower starts within 700~ns of the event start time and its energy is greater than 25 (50)~MeV in the barrel (end cap)
region of the EMC~\cite{BESIII}.
The energy deposited in the nearby TOF counters is included to improve the reconstruction efficiency and energy resolution. The opening angle between
the candidate shower and the nearest charged track must be greater than $10^{\circ}$. The $\gamma\gamma$ pair is taken as a $\pi^0$ candidate
if its invariant mass is within $(0.115,\,0.150)$\,GeV$/c^{2}$. To improve the resolution, a kinematic fit constraining the $\gamma\gamma$
invariant mass to the $\pi^{0}$ nominal mass~\cite{pdg2018} is imposed on the selected photon pair.

\section{Yields of ST $\bar D$ mesons}

To select $\bar D^0\to K^+\pi^-$ candidates, the backgrounds from cosmic rays and Bhabha events are rejected by using the same requirements described in Ref.~\cite{deltakpi}.
In the selection of $\bar D^0\to K^+\pi^-\pi^-\pi^+$ candidates, the $\bar D^0\to K^0_SK^\pm\pi^\mp$ decays are suppressed by requiring the mass of all $\pi^+\pi^-$ pairs to
be outside $(0.478,0.518)$~GeV/$c^2$.%, to simplify the BF measurements.

The tagged $\bar D$ mesons are identified using two variables, namely the energy difference
\begin{equation}
\Delta E_{\rm tag} \equiv E_{\rm tag} - E_{\rm b},
\label{eq:deltaE}
\end{equation}
and the beam-constrained mass
\begin{equation}
M_{\rm BC}^{\rm tag} \equiv \sqrt{E^{2}_{\rm b}-|\vec{p}_{\rm tag}|^{2}}.
\label{eq:mBC}
\end{equation}
Here, $E_{\rm b}$ is the beam energy,
$\vec{p}_{\rm tag}$ and $E_{\rm tag}$ are the momentum and energy of
the $\bar D$ candidate in the rest frame of $e^+e^-$ system, respectively.
For each tag mode, if there are multiple candidates in an event,
only the one with the smallest $|\Delta E_{\rm tag}|$ is kept.
The tagged $\bar D$ candidates are required to satisfy
$\Delta E_{\rm tag}\in(-55,40)$\,MeV for the tag modes containing $\pi^0$ in the final states
and $\Delta E_{\rm tag}\in(-25,25)$\,MeV for the other tag modes, thereby taking into account the different resolutions.

\begin{figure}[htp]
  \centering
\includegraphics[width=1.0\linewidth]{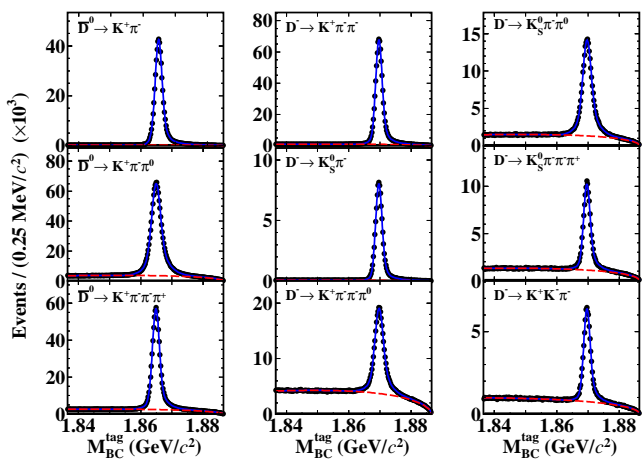}
  \caption{\small
Fits to the $M_{\rm BC}$ distributions of
the ST $\bar D^0$ (left column) and $D^-$ (middle and right columns) candidates,
where the points with error bars are data,
the blue solid and red dashed curves are the fit results
and the fitted backgrounds, respectively.}
\label{fig:datafit_MassBC}
\end{figure}

To extract the yields of ST $\bar D$ mesons for individual tag modes, binned-maximum likelihood fits are performed on the corresponding $M_{\rm BC}^{\rm tag}$
distributions of the accepted ST candidates following Refs.~\cite{epjc76,cpc40,bes3-pimuv,bes3-Dp-K1ev,bes3-etaetapi,bes3-omegamuv,bes3-etamuv}.
In the fits, the $\bar D$ signal is modeled by an MC-simulated shape convolved with
a double-Gaussian function describing the resolution difference between data and MC simulation.
The combinatorial background shape is described by an ARGUS function~\cite{ARGUS}
defined as $c_f(f;E_{\rm end},\xi_f)=A_f\cdot f\cdot \sqrt{1 - \frac {f^2}{E^2_{\rm end}}} \cdot \exp\left[\xi_f \left(1-\frac {f^2}{E^2_{\rm end}}\right)\right]$,
where $f$ denotes $M^{\rm tag}_{\rm BC}$, $E_{\rm end}$ is an endpoint fixed at 1.8865 GeV, $A_f$ is a normalization factor, and $\xi_f$ is a free parameter.
The resulting fits to the $M_{\rm BC}$
distributions for each mode are shown in
Fig.~\ref{fig:datafit_MassBC}.
The total yields of the ST $\bar D^0$  and $D^-$ mesons in data are $2327839\pm1860$ and
$1558159\pm2113$, respectively, where the uncertainties are statistical only.

\section{Yields of DT events}

In the recoiling sides against the tagged $\bar D$ candidates, the signal $D$ decays are selected
by using the residual tracks that have not been used to reconstruct the tagged $\bar D$ candidates.
To suppress the $K^0_S$ contribution in the individual mass spectra for the
$D^0\to K^+K^-\pi^0\pi^0$, $D^0\to K^0_SK^0_S\pi^{+}\pi^{-}$, and $D^+\to K^0_SK^+\pi^+\pi^-$ decays, the
$\pi^{+}\pi^{-}$ and $\pi^{0}\pi^{0}$ invariant masses are required to be outside $(0.468,0.528)$~GeV/$c^2$ and $(0.438,0.538)$~GeV/$c^2$, respectively.
To suppress the background from $D^0\to K^-\pi^+\omega$ in the identification of the $D^0\to K^0_SK^-\pi^+\pi^0$ process,
the $K^0_S\pi^0$ invariant mass is required to be outside $(0.742,0.822)$ GeV/$c^2$.
These requirements correspond to at least five times the fitted mass resolution away from the fitted mean of the mass peak.

The signal $D$ mesons are identified using the energy difference $\Delta E_{\rm sig}$
and the beam-constrained mass $M_{\rm BC}^{\rm sig}$, which are calculated with Eqs.~(\ref{eq:deltaE}) and (\ref{eq:mBC})
by substituting ``tag'' with ``sig''.
For each signal mode, if there are multiple candidates in an event, only the one with the smallest $|\Delta E_{\rm sig}|$ is kept.
The signal decays are required to satisfy the mode-dependent $\Delta E_{\rm sig}$ requirements,
as shown in the second column of Table~\ref{tab:DT}.
To suppress incorrectly identified $D\bar D$ candidates, the opening angle between the tagged $\bar D$ and the signal $D$ is required to be greater than $160^\circ$,
resulting in a loss of (2-6)\% of the signal and suppressing (8-55)\% of the background.
%[JM: SO WHAT IS SUPPRESSION FACTOR THEN IN THIS CONTEXT?]

\begin{figure}[htp]
  \centering
  \includegraphics[width=1.0\linewidth]{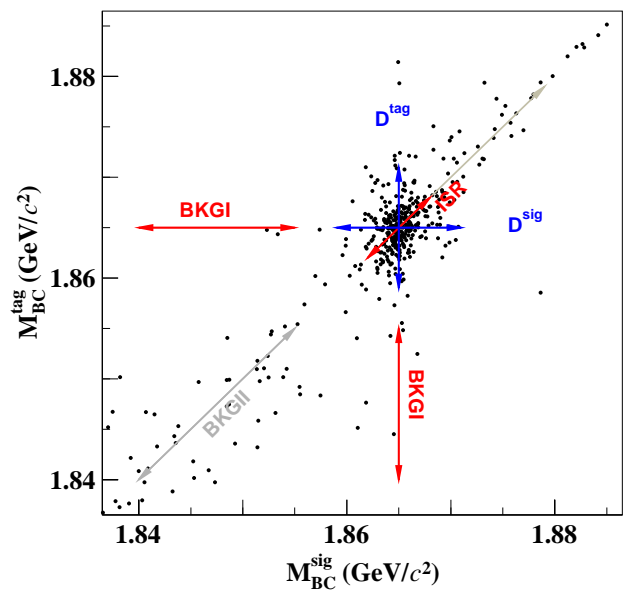}
  \caption{
    The $M_{\rm BC}^{\rm tag}$
    versus $M_{\rm BC}^{\rm sig}$ distribution of the accepted DT candidates of $D^+\to K^+K^-\pi^+\pi^0$ in data.
    % for example.
    Here, ISR denotes the signal spreading along the diagonal direction.
}
\label{fig:mBC2D}
\end{figure}

Figure~\ref{fig:mBC2D} shows the $M_{\rm BC}^{\rm tag}$
versus $M_{\rm BC}^{\rm sig}$ distribution of the accepted DT candidates in data.
The signal events concentrate around $M_{\rm BC}^{\rm tag} = M_{\rm BC}^{\rm sig} = M_{D}$,
where $M_{D}$ is the nominal $D$ mass~\cite{pdg2018}.
The events with correctly reconstructed $D$ ($\bar D$) and incorrectly
reconstructed $\bar D$ ($D$), named BKGI, are spread along the lines around
$M_{\rm BC}^{\rm tag} = M_{D}$ or $M_{\rm BC}^{\rm sig} = M_{D}$.
The events smeared along the diagonal, named BKGII,
are mainly from the $e^+e^- \to q\bar q$ processes.
The events with uncorrelated and incorrectly reconstructed $D$ and $\bar D$, named BKGIII,
disperse across the whole allowed kinematic region.

For each signal $D$ decay mode, the yield of DT events ($N^{\rm fit}_{\rm DT}$) is obtained from a two-dimensional (2D) unbinned maximum-likelihood
fit~\cite{cleo-2Dfit} on the $M_{\rm BC}^{\rm tag}$ versus $M_{\rm BC}^{\rm sig}$ distribution of the accepted candidates. In the fit, the probability
density functions (PDFs) of signal, BKGI, BKGII, and BKGIII are constructed as
\begin{itemize}
\item
signal: $a(x,y)$,
\item
BKGI: $b(x)\cdot c_y(y;E_{\rm b},\xi_{y}) + b(y)\cdot c_x(x;E_{\rm b},\xi_{x})$,
\item
BKGII: $c_z(z;\sqrt{2}E_{\rm b},\xi_{z}) \cdot g(k)$, and
\item
BKGIII: $c_x(x;E_{\rm b},\xi_{x}) \cdot c_y(y;E_{\rm b},\xi_{y})$,
\end{itemize}
respectively.
Here, $x=M_{\rm BC}^{\rm sig}$, $y=M_{\rm BC}^{\rm tag}$, $z=(x+y)/\sqrt{2}$, and $k=(x-y)/\sqrt{2}$.
The PDFs of signal $a(x,y)$, $b(x)$, and $b(y)$ are described by the corresponding MC-simulated shapes.
$c_f(f;E_{\rm end},\xi_f)$ is an ARGUS function~\cite{ARGUS} defined above,
where $f$ denotes $x$, $y$, or $z$; $E_{\rm b}$ is fixed at 1.8865 GeV.
$g(k)$ is a Gaussian function with mean of zero and standard deviation parametrized by $\sigma_k=\sigma_0 \cdot(\sqrt{2}E_{\rm b}/c^2-z)^p$,
where $\sigma_0$ and $p$ are fit parameters.

\begin{figure}[htp]
  \centering
  \includegraphics[width=1.0\linewidth]{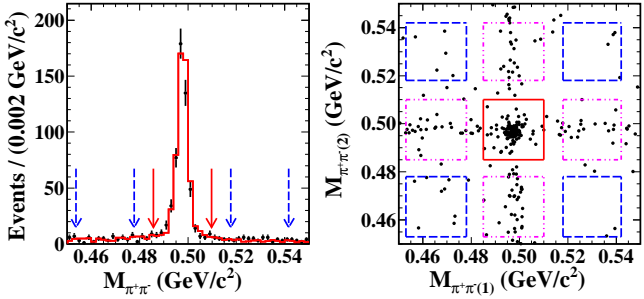}
\caption{\small
(a)~The $\pi^+\pi^-$ invariant-mass distributions of the $D^+\to K^0_SK^-\pi^+\pi^+$ candidate events
of data (points with error bars) and inclusive MC sample (histogram).
Pairs of the red solid~(blue dashed) arrows denote the $K^0_S$ signal~(sideband) regions.
(b)~Distribution of $M_{\pi^+\pi^-(1)}$ versus $M_{\pi^+\pi^-(2)}$ for the $D^0\to K^0_SK^0_S\pi^+\pi^-$ candidate events in data.
Red solid box denotes the 2D signal region.
Pink dot-dashed~(blue dashed) boxes indicate the 2D sideband 1~(2) regions.
}\label{fig:mks}
\end{figure}

Combinatorial $\pi^+\pi^-$ pairs from the decays
$D^0\to K^0_S2(\pi^+\pi^-)$ [and $D^0\to 3(\pi^+\pi^-)$], $D^0\to K^-\pi^+\pi^+\pi^-\pi^0$, $D^0\to K^+\pi^+\pi^-\pi^-\pi^0$, $D^+\to K^-\pi^+\pi^+\pi^+\pi^-$, $D^+\to K^+2(\pi^+\pi^-)$, $D^+\to K^+\pi^+\pi^-\pi^0\pi^0$, $D^+\to K^0_S\pi^+\pi^+\pi^-\pi^0$ [and $D^+\to 2(\pi^+\pi^-)\pi^+\pi^0$]
may also satisfy the $K^0_S$ selection criteria
and form peaking backgrounds around $M_D$ in the $M_{\rm BC}^{\rm sig}$ distributions for $D^0\to K^0_SK^0_S\pi^+\pi^-$,
$D^0\to K^0_SK^-\pi^+\pi^0$,
$D^0\to K^0_SK^+\pi^-\pi^0$,
$D^+\to K^0_SK^+\pi^0\pi^0$
$D^+\to K^0_SK^-\pi^+\pi^+$,
$D^+\to K^0_SK^+\pi^+\pi^-$, and
$D^+\to K^0_SK^0_S\pi^+\pi^0$, respectively.
This kind of peaking background is estimated by selecting events in the $K^0_S$ sideband region of
$(0.454,0.478)\cup(0.518,0.542)~{\rm GeV}/c^2$.
For $D^0\to K^0_SK^-\pi^+\pi^0$, $D^0\to K^0_SK^+\pi^-\pi^0$, $D^+\to K^0_SK^-\pi^+\pi^+$, $D^+\to K^0_SK^+\pi^+\pi^-$, and $D^+\to K^0_SK^+\pi^0\pi^0$ decays,
one-dimensional (1D) signal and sideband regions are used.
For $D^0\to K^0_SK^0_S\pi^+\pi^-$ and $D^+\to K^0_SK^0_S\pi^+\pi^0$ decays, 2D signal and sideband regions are used.
The 2D $K^0_S$ signal region is defined as the square region with both $\pi^+\pi^-$ combinations lying in the $K^0_S$ signal regions.
The 2D $K^0_S$ sideband 1~(2) regions are defined as the square regions with 1~(2) $\pi^+\pi^-$ combination(s) located in the 1D $K^0_S$ sideband regions
and the rest in the 1D $K^0_S$ signal region.
Figure~\ref{fig:mks} shows 1D and 2D $\pi^+\pi^-$ invariant-mass distributions as well as the $K^0_S$ signal and sideband regions.

\begin{figure*}[htbp]
  \centering
\includegraphics[width=0.49\linewidth]{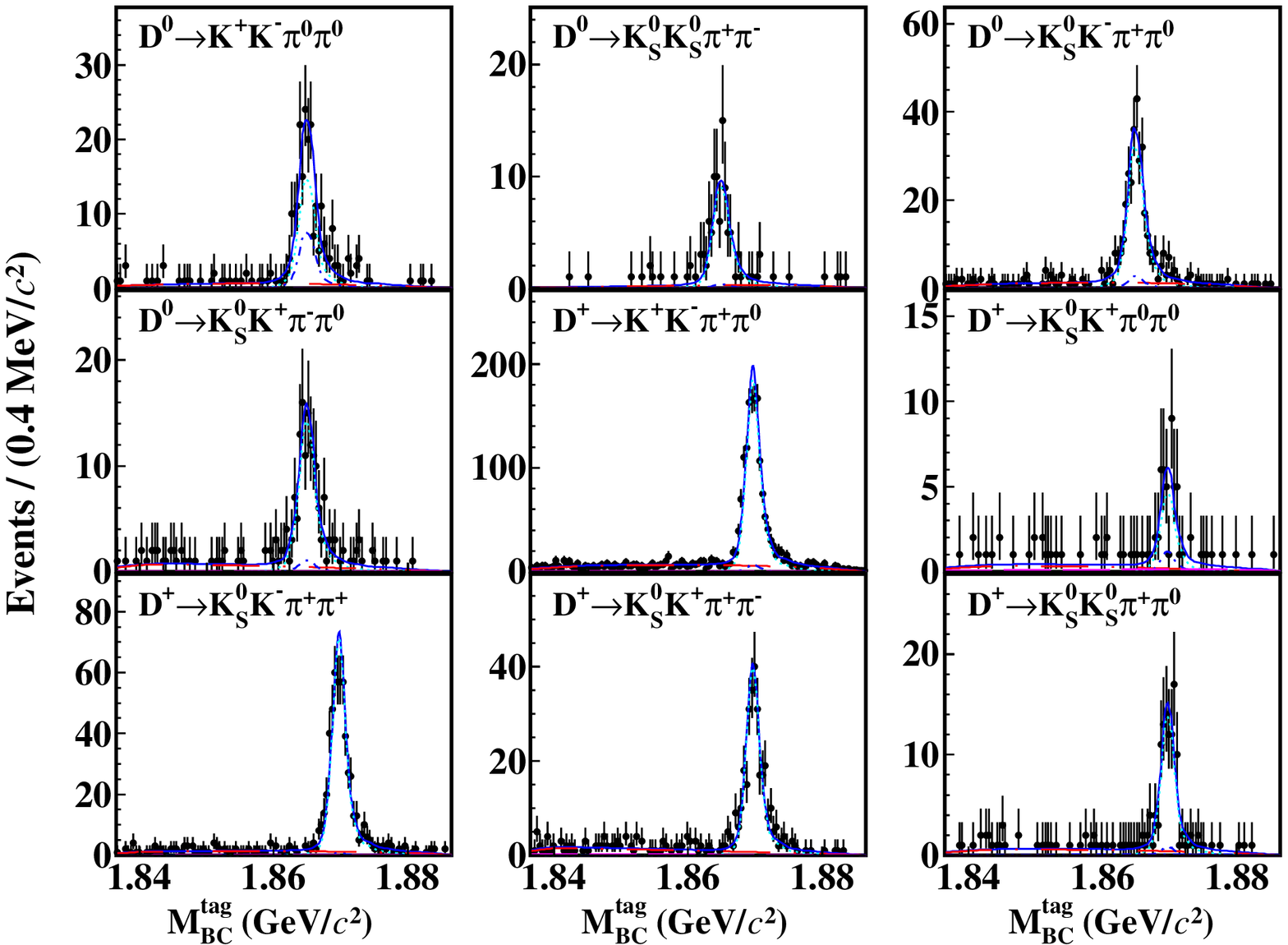}
\includegraphics[width=0.49\linewidth]{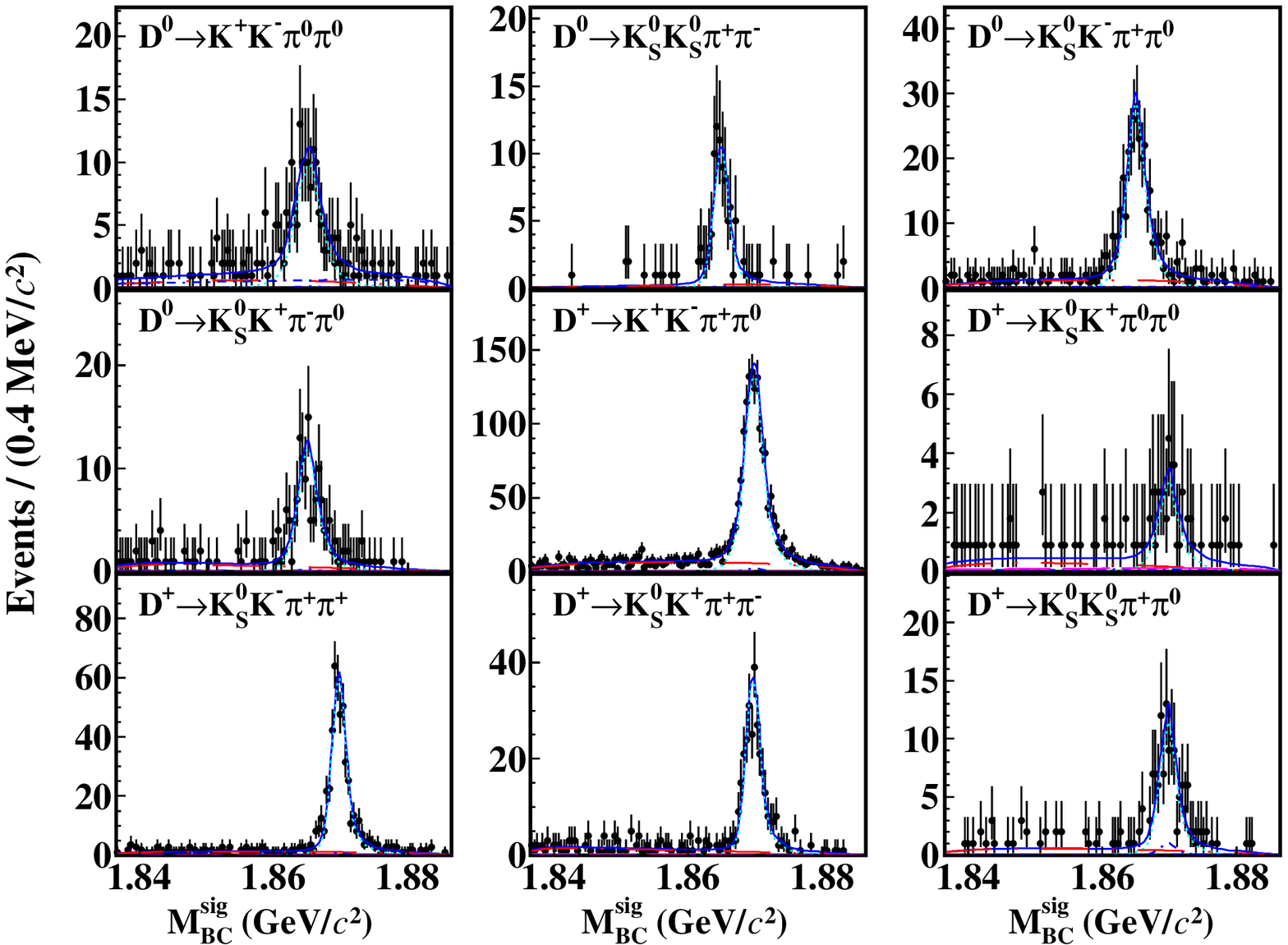}
  \caption{\small
Projections on the $M^{\rm tag}_{\rm BC}$ and
$M^{\rm sig}_{\rm BC}$ distributions of the 2D fits to the DT candidate events with all $\bar D^0$ or $D^-$ tags.
Data are shown as points with error bars.
Blue solid, light blue dotted, blue dot-dashed, red dot-long-dashed,
and pink long-dashed curves denote the overall fit results,
signal, BKGI, BKGII, and BKGIII components (see text), respectively.
}
\label{fig:2Dfit}
\end{figure*}

For the signal decays involving $K^0_S$ meson(s) in the final states,
the net yields of DT events are calculated by subtracting the sideband contribution from the DT fitted yield by
\begin{equation}
\label{eq:1}
N^{\rm net}_{\rm DT} = N^{\rm fit}_{\rm DT} + \Sigma^N_i \left [\left (-\frac{1}{2} \right )^i N^{\rm fit}_{{\rm sid}i} \right ].
\end{equation}
Here, $N=1$ for the decays with one $K^0_S$ meson while $N=2$ for the decays with two $K^0_S$ mesons.
The combinatorial $\pi^+\pi^-$ backgrounds are assumed to be uniformly distributed and double-counting is avoided by subtracting (2) yields from (1)
yields appropriately.
$N^{\rm fit}_{\rm DT}$ and $N^{\rm fit}_{{\rm sid}i}$
are the fitted $D$ yields in the 1D or 2D signal region and sideband $i$ region, respectively.
For the other signal decays, the net yields of DT events are $N^{\rm fit}_{\rm DT}$. Figure~\ref{fig:2Dfit} shows the $M^{\rm tag}_{\rm BC}$ and
$M^{\rm sig}_{\rm BC}$ projections of the 2D fits to data. From these 2D fits, we obtain the DT yields for individual signal decays as shown in Table~\ref{tab:DT}.

For each signal decay mode, the statistical significance is calculated according to
$\sqrt{-2{\rm ln ({\mathcal L_0}/{\mathcal L_{\rm max}}})}$,
where ${\mathcal L}_{\rm max}$ and ${\mathcal L}_0$ are the maximum likelihoods of the fits with and without involving the signal component, respectively.
The effect of combinatorial $\pi^+\pi^-$ backgrounds in the $K^0_S$-signal regions has been considered for the decays involving a $K^0_S$.
The statistical significance for each signal decay is found to be greater than $8\sigma$.

\section{Results}

\begin{figure*}[htbp]
  \centering
\includegraphics[width=0.8\linewidth]{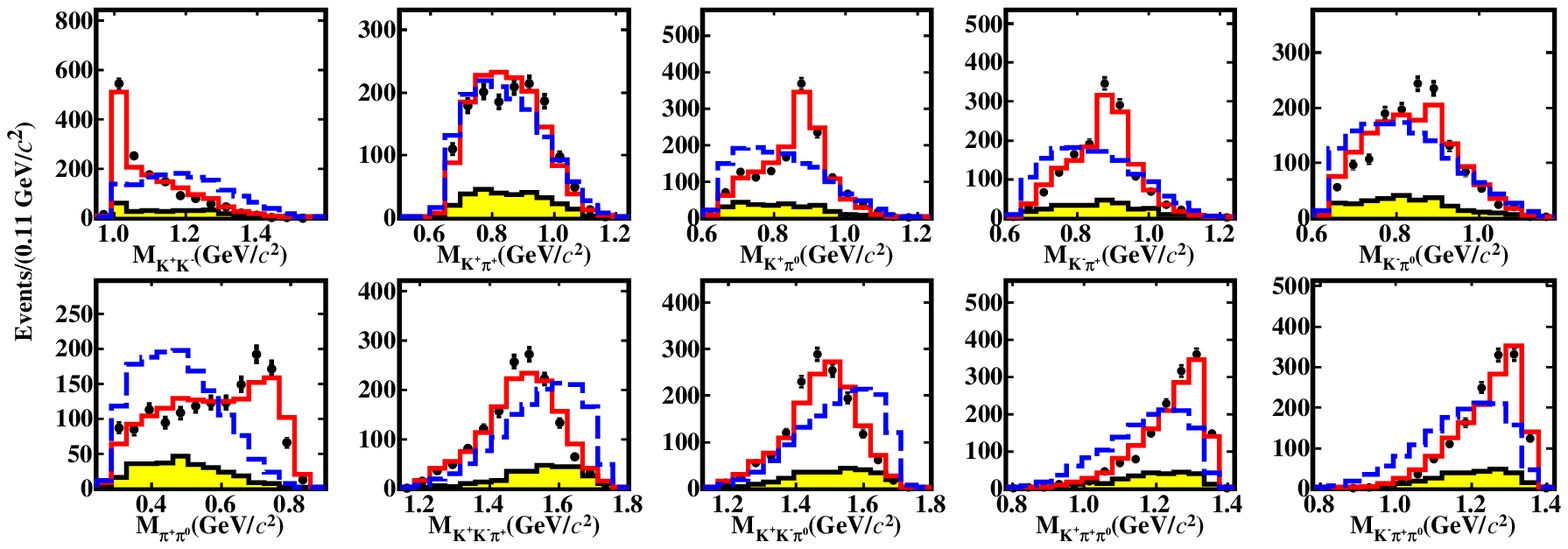}
  \caption{\small
The invariant mass distributions of two or three-body particle combinations of $D^+\to K^+K^-\pi^+\pi^0$ candidate events for data and MC simulations.
Data are shown as points with error bars. Red solid histograms are mixed signal MC samples. Blue dashed histograms are PHSP signal MC samples. Yellow hatched histograms are the backgrounds estimated from the inclusive MC sample.
}
\label{add}
\end{figure*}

Each of the $D^0\to K^0_SK^-\pi^+\pi^0$, $D^+\to K^+K^-\pi^+\pi^0$, $D^+\to K^0_SK^-\pi^+\pi^+$, and $D^+\to K^0_SK^+\pi^+\pi^-$ decays
is modeled by the corresponding mixed signal MC samples, in which the dominant decay modes containing resonances of
$K^*(892)$, $\rho(770)$, and $\phi$ are mixed with the phase space (PHSP) signal MC samples. The mixing ratios are determined by
examining the corresponding invariant mass and momentum spectra. The other decays, which are limited in statistics, are generated with the PHSP generator.
The momentum and the polar angle distributions of the daughter particles and the invariant masses of each two- and three-body particle combinations
of the data agree with those of the MC simulations. As an example, Fig.~\ref{add} shows the invariant mass distributions of two or three-body particle combinations of $D^+\to K^+K^-\pi^+\pi^0$ candidate events for data and MC simulations.

The measured values of $N^{\rm net}_{{\rm DT}}$,
$\epsilon^{}_{{\rm sig}}$, and the obtained BFs are summarized in Table~\ref{tab:DT}. The current world-average values are also given for comparison.
The signal efficiencies have been corrected by the necessary data-MC differences in the selection efficiencies of $K^\pm$ and $\pi^\pm$ tracking and PID
procedures and the $\pi^0$ reconstruction.
%(as well as the quantum correlation factor for neutral $D$ decays, as described later).
These efficiencies also include the BFs of the $K^0_S$ and $\pi^0$ decays.
The efficiency for $D^+\to K^0_SK^+\pi^+\pi^-$ ($D^0\to K^0_SK^-\pi^+\pi^0$) is lower than that of $D^+\to K^0_SK^-\pi^+\pi^+$ ($D^0\to K^0_SK^+\pi^-\pi^0$)
due to the $K^0_S$ $(\omega)$ rejection in the $\pi^+\pi^-$ ($K^0_S\pi^0$) mass spectrum.

\section{Systematic uncertainties}

The systematic uncertainties are estimated relative to the measured BFs and are discussed below.
In BF determinations using Eq.~(\ref{eq:br}), all uncertainties associated with the selection of tagged $\bar D$ canceled in the ratio.
The systematic uncertainties in the total yields of ST $\bar D$ mesons related to the $M_{\rm BC}$ fits to the ST
$\bar D$ candidates, were previously estimated to be
0.5\% for both neutral and charged $\bar D$~\cite{epjc76,cpc40,bes3-pimuv}.

The tracking and PID efficiencies for $K^\pm$ or $\pi^\pm$, $\epsilon_{K\,{\rm or}\,\pi}^{\rm tracking\,(PID)}[{\rm data}]$ and $\epsilon_{K\,{\rm or}\,\pi}^{\rm tracking\,(PID)}[{\rm MC}]$,
are investigated using DT $D\bar D$ hadronic events.
The averaged ratios between data and MC efficiencies ($f_{K\,{\rm or}\,\pi}^{\rm tracking\,(PID)}=\epsilon_{K\,{\rm or}\,\pi}^{\rm tracking\,(PID)}[{\rm data}]/\epsilon_{K\,{\rm or}\,\pi}^{\rm tracking\,(PID)}[{\rm MC}]$) of tracking (PID) for $K^\pm$ or $\pi^\pm$ are weighted by the corresponding momentum spectra of signal MC events,
giving $f_K^{\rm tracking}$ to be $1.022{\text -}1.031$ and $f_\pi^{\rm tracking}$ to be close to unity. After correcting the MC efficiencies by $f_K^{\rm tracking}$,
the residual uncertainties of $f_{K\,{\rm or}\,\pi}^{\rm tracking}$ are assigned as the systematic uncertainties of tracking efficiencies,
which are (0.4-0.7)\% per $K^\pm$ and (0.2-0.3)\% per $\pi^\pm$. $f_K^{\rm PID}$ and $f_\pi^{\rm PID}$ are all close to unity and their individual uncertainties, (0.2-0.3)\%,
are taken as the associated systematic uncertainties per $K^\pm$ or $\pi^\pm$.

The systematic error related to the uncertainty in the $K_{S}^{0}$ reconstruction efficiency
is estimated from measurements of $J/\psi\to K^{*}(892)^{\mp}K^{\pm}$ and $J/\psi\to \phi K_S^{0}K^{\pm}\pi^{\mp}$ control samples~\cite{sysks}
and found to be 1.6\% per $K^0_S$.
The systematic uncertainty of $\pi^0$ reconstruction efficiency is assigned as (0.7-0.8)\% per $\pi^0$
from a study of DT $D\bar D$ hadronic decays of
$\bar D^0\to K^+\pi^-\pi^0$ and $\bar D^0\to K^0_S\pi^0$ decays tagged by either $D^0\to K^-\pi^+$ or $D^0\to K^-\pi^+\pi^+\pi^-$~\cite{epjc76,cpc40}.

The systematic uncertainty in the 2D fit to the $M_{\rm BC}^{\rm tag}$ versus $M_{\rm BC}^{\rm sig}$ distribution is examined via the repeated measurements in which the signal shape
and the endpoint of the ARGUS function ($\pm0.2$\,MeV/$c^2$) are varied.
Quadratically summing the changes of the BFs for these two sources yields the corresponding systematic uncertainties.

The systematic uncertainty due to the $\Delta E_{\rm sig}$ requirement is assigned to be 0.3\%,
which corresponds to the largest efficiency difference with and without smearing the data-MC Gaussian resolution of $\Delta E_{\rm sig}$ for signal MC events.
Here, the smeared Gaussian parameters are obtained by using the samples of DT events $D^0\to K^0_S\pi^0$, $D^0\to K^-\pi^+\pi^0$, $D^0\to K^-\pi^+\pi^0\pi^0$, and $D^+\to K^-\pi^+\pi^+\pi^0$ versus the same $\bar D$ tags in our nominal analysis.
The systematic uncertainties due to $K^0_S$ sideband choice and $K^0_S$ rejection mass window
are assigned
by examining the changes of the BFs via varying nominal $K^0_S$ sideband and corresponding rejection window by $\pm5$~MeV/$c^2$.

For the decays whose efficiencies are estimated with mixed signal MC events,
the systematic uncertainty in the MC modeling is determined by comparing the signal efficiency when changing the percentage of MC sample components.
For the decays whose efficiencies are estimated with PHSP-distributed signal MC events, the uncertainties are assigned as the change of the signal efficiency after adding the possible decays containing $K^*(892)$ or $\rho(770)$.
The imperfect simulations of the momentum and $\cos\theta$ distributions of charged particles are considered as a source of systematic uncertainty.
The signal efficiencies are re-weighted by those distributions in data with background subtracted.
The largest change of the re-weighted to nominal efficiencies, 0.9\%, is assigned as the corresponding systematic uncertainty.

The measurements of the BFs of the neutral $D$ decays are affected by quantum correlation effect. For each neutral $D$ decay, the $CP$-even component is estimated by the $CP$-even tag $D^0\to K^+K^-$
and the $CP$-odd tag $D^0\to K^0_S\pi^0$.
Using the same method as described in Ref.~\cite{QC-factor}
and the necessary parameters quoted from Refs.~\cite{R-ref1,R-ref2,R-ref3},
we find the correction factors to account for the quantum correlation effect on the measured BFs are
$(98.3^{+1.6}_{-1.1{\,\rm stat}})\%$, $(98.1^{+2.8}_{-1.7{\,\rm stat}})\%$, $(95.9^{+3.4}_{-2.7{\,\rm stat}})\%$, and
$(98.4^{+1.1}_{-1.0{\,\rm stat}})\%$ for $D^0\to K^+K^-\pi^0\pi^0$, $D^0\to K^0_SK^0_S\pi^+\pi^-$,
$D^0\to K^0_SK^-\pi^+\pi^0$, and $D^0\to K^0_SK^+\pi^-\pi^0$, respectively.
After correcting the signal efficiencies by the individual factors, the residual uncertainties are
assigned as systematic uncertainties.

The uncertainties due to the limited MC statistics for various signal decays, (0.4-0.8)\%, are taken into account as a systematic uncertainty.
The uncertainties of the quoted BFs of the $K^0_S\to \pi^+\pi^-$ and
$\pi^0\to \gamma\gamma$ decays are 0.07\% and 0.03\%, respectively~\cite{pdg2018}.

The efficiencies of $D\bar D$ opening angle requirement is studied by using the DT events of $D^0\to K^-\pi^+\pi^+\pi^-$, $D^0\to K^-\pi^+\pi^0\pi^0$, and $D^+\to K^-\pi^+\pi^+\pi^0$ tagged by the same tag modes in our nominal analysis. The difference of the accepted efficiencies between data and MC simulations,
0.4\% for the decays without $\pi^0$,
0.8\% for the decays involving one $\pi^0$ and
0.3\% for the decays involving two $\pi^0$s,
is assigned as the associated systematic uncertainty.

Table~\ref{tab:relsysuncertainties1} summarizes the systematic uncertainties in the BF measurements. For each signal decay,
the total systematic uncertainty is obtained by adding the above
effects in quadrature to be (2.6-6.0)\% for various signal decay modes.

\begin{table*}[htbp]
\centering
\caption{\small
Requirements of $\Delta E_{\rm sig}$,
net yields of DT candidates ($N^{\rm net}_{{\rm DT}}$),
signal efficiencies ($\epsilon_{\rm sig}$),
and the obtained BFs (${\mathcal B}_{\rm sig}$) for various signal decays as well as comparisons with the world-average BFs (${\mathcal B}_{\rm PDG}$).
The first and second uncertainties for ${\mathcal B}_{\rm sig}$ are statistical and systematic, respectively,
while the uncertainties for $N^{\rm net}_{\rm DT}$ and $\epsilon_{\rm sig}$ are statistical only.
The world-average BF of $D^+\to K^+K^-\pi^+\pi^0$ is obtained by summing over the contributions of $D^+\to \phi(\to K^+K^-)\pi^+\pi^0$ and $D^+\to K^+K^-\pi^+\pi^0|_{{\rm non\text-}\phi}$.
}\label{tab:DT}
\begin{ruledtabular}
\begin{tabular}{lcccccc}
\multicolumn{1}{c} {Signal mode}&$\Delta E_{\rm sig}$\,(MeV) &$N^{\rm net}_{\rm DT}$ & $\epsilon_{\rm sig}$\,(\%) &
  ${\mathcal B}_{\rm sig}$\,($\times10^{-3}$) & ${\mathcal B}_{\rm PDG}$\,($\times10^{-3}$) \\  \hline
$D^0\to K^+K^-\pi^0\pi^0$    &$(-59,40)$&$ 132.1\pm13.9$&$ 8.20\pm0.07$&$0.69\pm0.07\pm0.04$&--\\
$D^0\to K^0_SK^0_S\pi^+\pi^-$&$(-22,22)$&$  62.5\pm10.4$&$ 5.14\pm0.04$&$0.52\pm0.09\pm0.03$&$1.22\pm0.23$\\
$D^0\to K^0_SK^-\pi^+\pi^0$  &$(-43,32)$&$ 195.8\pm20.3$&$ 6.38\pm0.06$&$1.32\pm0.14\pm0.07$&--\\
$D^0\to K^0_SK^+\pi^-\pi^0$  &$(-44,33)$&$ 119.3\pm12.9$&$ 7.94\pm0.06$&$0.65\pm0.07\pm0.02$&--\\
$D^+\to K^+K^-\pi^+\pi^0$    &$(-39,30)$&$1311.7\pm40.4$&$12.72\pm0.08$&$6.62\pm0.20\pm0.25$&$26^{+9}_{-8}$\\
$D^+\to K^0_SK^+\pi^0\pi^0$  &$(-61,44)$&$  34.7\pm 7.2$&$ 3.77\pm0.02$&$0.59\pm0.12\pm0.04$&--\\
$D^+\to K^0_SK^-\pi^+\pi^+$  &$(-22,21)$&$ 467.9\pm26.6$&$13.24\pm0.08$&$2.27\pm0.12\pm0.06$&$2.38\pm0.17$\\
$D^+\to K^0_SK^+\pi^+\pi^-$  &$(-21,20)$&$ 279.6\pm18.1$&$ 9.39\pm0.06$&$1.91\pm0.12\pm0.05$&$1.74\pm0.18$\\
$D^+\to K^0_SK^0_S\pi^+\pi^0$&$(-46,37)$&$  80.4\pm12.0$&$ 3.84\pm0.03$&$1.34\pm0.20\pm0.06$&--\\
\end{tabular}
\end{ruledtabular}
\end{table*}

\begin{table*}[htp]
\centering
\caption{
Systematic uncertainties (\%) in the measurements of the BFs of the signal decays
(1) $D^0\to K^+K^-\pi^0\pi^0$,
(2) $D^0\to K^0_SK^0_S\pi^+\pi^-$,
(3) $D^0\to K^0_SK^-\pi^+\pi^0$,
(4) $D^0\to K^0_SK^+\pi^-\pi^0$,
(5) $D^+\to K^+K^-\pi^+\pi^0$,
(6) $D^+\to K^0_SK^+\pi^0\pi^0$,
(7) $D^+\to K^0_SK^-\pi^+\pi^+$,
(8) $D^+\to K^0_SK^+\pi^+\pi^-$, and
(9) $D^+\to K^0_SK^0_S\pi^+\pi^0$.}
\label{tab:relsysuncertainties1}
\centering
  \begin{ruledtabular}
\begin{tabular}{cccccccccc}
Source/Signal decay              & 1  & 2  &  3 & 4  & 5  & 6  & 7  & 8  & 9   \\ \hline
$N^{\rm tot}_{\rm ST}$           &0.5 &0.5 &0.5 &0.5 &0.5 &0.5 &0.5 &0.5 &0.5  \\
$(K/\pi)^\pm$ tracking           &1.0 &0.6 &0.9 &0.9 &1.6 &0.4 &1.1 &1.2 &0.3  \\
$(K/\pi)^\pm$ PID                &0.4 &0.4 &0.6 &0.6 &1.0 &0.2 &0.6 &0.7 &0.2  \\
$K^0_S$ reconstruction           &... &3.2 &1.6 &1.6 &... &1.6 &1.6 &1.6 &3.2  \\
$\pi^0$ reconstruction           &1.6 &... &0.7 &0.7 &0.8 &1.6 &... &... &0.7  \\
$\Delta E_{\rm sig}$ requirement &0.7 &0.7 &0.7 &0.7 &0.7 &0.7 &0.7 &0.7 &0.7  \\
$K_{S}^{0}$ rejection            &4.2 &2.4 &... &... &... &4.2 &... &0.8 &...  \\
$K_{S}^{0}$ sideband             &... &0.2 &1.1 &0.2 &... &1.3 &0.1 &0.1 &0.2  \\
Quoted BFs                       &0.0 &0.1 &0.1 &0.1 &0.0 &0.1 &0.1 &0.1 &0.1  \\
MC statistics                    &0.8 &0.6 &0.7 &0.6 &0.5 &0.4 &0.4 &0.5 &0.6  \\
MC modeling                      &1.3 &1.0 &0.5 &0.7 &2.1 &1.4 &0.5 &0.7 &0.5  \\
Imperfect simulation             &0.9 &0.9 &0.9 &0.9 &0.9 &0.9 &0.9 &0.9 &0.9  \\
$D\bar D$ opening angle          &0.3 &0.4 &0.8 &0.8 &0.8 &0.3 &0.4 &0.4 &0.8  \\
2D fit                           &1.3 &2.8 &3.1 &1.5 &1.9 &2.7 &0.5 &0.6 &3.0  \\
Quantum correlation effect       &1.6 &2.8 &3.4 &1.1 &... &... &... &... &...  \\
\hline
Total                            &5.5 &5.9 &5.4 &3.3 &3.8 &6.0 &2.6 &2.8 &4.8  \\
\end{tabular}
  \end{ruledtabular}
\end{table*}

\section{Summary}

In summary, by analyzing a data sample obtained in $e^+e^-$ collisions at $\sqrt{s}=3.773$~GeV with the BESIII detector and
corresponding to an integrated luminosity of 2.93~fb$^{-1}$, we obtained the first direct measurements of the absolute BFs of
nine $D^{0(+)}\to K\bar K\pi\pi$ decays containing $K^0_S$ or $\pi^0$ mesons.
The $D^0\to K^+K^-\pi^0\pi^0$, $D^0\to K^0_SK^-\pi^+\pi^0$, $D^0\to K^0_SK^+\pi^-\pi^0$, $D^+\to K^0_SK^+\pi^0\pi^0$,
and $D^+\to K^0_SK^0_S\pi^+\pi^0$ decays are observed for the first time. Compared to the world-average values, the BFs of
the $D^0\to K^0_SK^0_S\pi^+\pi^-$, $D^+\to K^+K^-\pi^+\pi^0$, $D^+\to K^0_SK^-\pi^+\pi^+$, and $D^+\to K^0_SK^+\pi^+\pi^-$ decays are measured with
improved precision. Our BFs of $D^+\to K^0_SK^-\pi^+\pi^+$ and $D^+\to K^0_SK^+\pi^+\pi^-$ are in agreement with individual world averages within $1\sigma$
while our BFs of $D^0\to K^0_SK^0_S\pi^+\pi^-$ and $D^+\to K^+K^-\pi^+\pi^0$
deviate with individual world averages by $2.3\sigma$ and $2.8\sigma$, respectively. The precision of the BF of $D^+\to K^+K^-\pi^+\pi^0$ is improved by a factor of
about seven.
Future amplitude analyses of all these $D^{0(+)}\to K\bar K\pi\pi$ decays with larger data samples foreseen at BESIII~\cite{bes3-white-paper}, Belle~II~\cite{belle2-white-paper}, and LHCb~\cite{lhcb-white-paper} will supply rich information of the two-body decay modes containing scalar, vector, axial and tensor mesons, thereby benefiting the understanding of quark SU(3)-flavor symmetry.

\section{Acknowledgement}

Authors thank for valuable discussions with Prof. Fu-sheng Yu.
The BESIII collaboration thanks the staff of BEPCII and the IHEP computing center for their strong support. This work is supported in part by National Key Basic Research Program of China under Contract No. 2015CB856700; National Natural Science Foundation of China (NSFC) under Contracts Nos.~11775230, 11475123, 11625523, 11635010, 11735014, 11822506, 11835012, 11935015, 11935016, 11935018, 11961141012; the Chinese Academy of Sciences (CAS) Large-Scale Scientific Facility Program; Joint Large-Scale Scientific Facility Funds of the NSFC and CAS under Contracts Nos.~U1532101, U1932102, U1732263, U1832207; CAS Key Research Program of Frontier Sciences under Contracts Nos. QYZDJ-SSW-SLH003, QYZDJ-SSW-SLH040; 100 Talents Program of CAS; INPAC and Shanghai Key Laboratory for Particle Physics and Cosmology; ERC under Contract No. 758462; German Research Foundation DFG under Contracts Nos. Collaborative Research Center CRC 1044, FOR 2359; Istituto Nazionale di Fisica Nucleare, Italy; Ministry of Development of Turkey under Contract No. DPT2006K-120470; National Science and Technology fund; STFC (United Kingdom); The Knut and Alice Wallenberg Foundation (Sweden) under Contract No. 2016.0157; The Royal Society, UK under Contracts Nos. DH140054, DH160214; The Swedish Research Council; U. S. Department of Energy under Contracts Nos. DE-FG02-05ER41374, DE-SC-0012069.

\end{document}